\documentclass[showpacs, twocolumn]{revtex4}

\usepackage{graphicx,textcomp}
\usepackage{dcolumn}
\usepackage{bm}

\begin{document}

\preprint{APS/123-QED}

\title{Alkali-helium snowball complexes formed on helium nanodroplets}
\author{S. M\"uller}
\author{M. Mudrich\email{mudrich@physik.uni-freiburg.de}}
\author{F. Stienkemeier}
\affiliation{Physikalisches Institut, Universit\"at Freiburg, 79104 Freiburg, Germany}
\date{\today}
%
\begin{abstract}
\noindent
We systematically investigate the formation and stability of snowballs formed by femtosecond photo-ionization of small alkali clusters bound to helium nanodroplets. For all studied alkali species Ak=(Na,\,K,\,Rb,\,Cs) we observe the formation of snowballs Ak$^+$He$_\mathrm{N}$ when multiply doping the droplets. Fragmentation of clusters Ak$_\mathrm{N}$ upon ionization appears to enhance snowball formation. In the case of Na and Cs we also detect snowballs Ak$_2^+$He$_\mathrm{N}$ formed around Ak dimer ions. While the snowball progression for Na and K is limited to less than 11\,helium atoms, the heavier atoms Rb and Cs feature wide distributions at least up to Ak$^+$He$_{41}$. Characteristic steps in the mass spectra of Cs-doped helium droplets are found at positions consistent with predictions on the closure of the 1$^\mathrm{st}$ shell of helium atoms around the Ak$^+$ ion based on variational Monte Carlo simulations.
\end{abstract}


\maketitle

\section{Introduction}

Ion-induced electrostatic interactions are of fundamental importance in various fields of physics, chemistry and biology~\cite{Yu:2003,Mordasini:2003}. Ions implanted in an ultracold helium (He) environment are of particular interest due to the influence of superfluidity on the mobility and on the He density distributions around the embedded ions. While electron impurities reside in bubble states with low He density due to repulsive interaction, positive ions induce local enhancement of the He density by electrostriction. Since the He density in this compressed shell surrounding the ion can exceed the density of solid He, such a structure is called a \textit{snowball}.

Using a phenomenological model by Atkins, the diameter of such a snowball in liquid He was estimated to about 6.5\,\AA\,consisting of about 50 atoms in two He layers, independent of the specific element~\cite{Atkins:1959}. For alkali and alkaline earth metals a distinct dependence on the specific element was observed in ion mobility measurements~\cite{Foerste:1997}. Compared to He$^+$, mobilities were found to be lower for alkali and higher for alkaline earth atoms, indicating larger, respectively lower, numbers of He atoms attached to the ions. In addition, those experiments showed evidence that mobility increases with the mass of the dopant~\cite{Foerste:1997}.

More recently, the stability of ion snowballs has been studied by means of abundance spectroscopy of snowballs formed upon ionization of metal atoms in He droplets~\cite{Doeppner:2001,Doeppner:2007,Tiggesbaumker:2007}. Snowballs with up to 150 He atoms attached to, e.\,g., Ag$^+$ and Mg$^+$ ions have been observed. Element specific stability patterns reflecting closures of He solvation shells are directly observable as steps or kinks in the snowball abundance distributions. For example, closures of the first and second solvation shells around positive silver ions, Ag$^+$, at N$_1$=10 and N$_2$=32 He atoms, respectively, are clearly visible as steps in the mass spectrum after femtosecond (fs) photo-ionization of Ag clusters embedded in He droplets. A sensitive dependence of the abundance of snowball complexes on the ionizing laser intensity is observed. Upon strong field ionization followed by massive fragmentation of the metal kernel, snowballs with as many as 150 (20) He atoms attached to the ionic Ag$^+$ (Mg$^+$) core appear in the mass spectra~\cite{Doeppner:2001,Doeppner:2007}. Moreover, the pump-probe dynamics of snowball formation was used as a diagnostic tool to investigate the caging of metal clusters in large He droplets~\cite{Doeppner:2007}.

A number of theoretical techniques have been applied to studying the solvation of positive ions in He droplets~\cite{Rossi:2004,Galli:2001,Nakayama:2000}. Mainly alkali and alkaline earth ions have been addressed so far since reliable Me$^+$--He potentials are available for these species. Using variational Monte Carlo simulations, Reatto and coworkers find that all alkali and alkaline-earth cations form snowball structures featuring shells of He atoms with high average density~\cite{Rossi:2004,Galli:2001,Buzzacchi:2001}. In addition to a modulated radial density profile around the impurity ions, snowballs are characterized by angular correlations in the first He shell as well as by the degree of radial localization of He atoms and the ability of He atoms to move from the first to the second shell and vice versa.

This solid-like order in the first shell turns out to be specific to the impurity ion. In contrast to the results of ion mobility measurements mentioned above, the number of He atoms bound to the impurity ions in the first shell, N$_1$, increases with larger ion mass. E.\,g., for the alkali ions Na$^+$, K$^+$, and Cs$^+$, Rossi \textit{et al.} have found N$_1$=12, 15 and 17.5, respectively~\cite{Rossi:2004}. Most pronounced local order is found for the Na$^+$ ion which is explained by the deep Na$^+$--He potential well and also by the matching of the minimum positions of the He--He and Na$^+$--He potentials. The calculated structure is compatible only with an icosahedron having 12 He atoms in the first shell. The Na$^+$He$_\mathrm{N}$ system was also studied by Nakayama and Yamashita~\cite{Nakayama:2000}. They predicted a triple layered structure with a rigid-body and solid-like first shell around the cation with 16 tightly bound He atoms in the first shell, considerably more than the 12 He atoms found in Ref.~\cite{Rossi:2004}.

The geometric structure of He shells in alkali-ion doped clusters was studied by quantum variational and diffusion Monte Carlo calculations~\cite{Marinetti:2007,Coccia:2007JCP,Coccia:2007WSPC}. By subsequently adding He atoms to the ion and computing the resulting cluster structure, marked differences in the Ak$^+$He$_\mathrm{N}$ geometry depending on N are found~\cite{Coccia:2007JCP}. The first solvation shell consists of equivalently arranged He atoms, i.e. their binding energy is determined by the two-body Ak$^+$--He potential alone, independently from the other solvent atoms. Due to the screening of the charge by the He atoms in the first shell, the further atoms are bound only by the He--He interaction. During buildup of the first shell different regular structures are observed to emerge for Li$^+$He$_6$ (octahedral), Li$^+$He$_8$ and Li$^+$He$_{10}$, Na$^+$He$_{10}$ and Na$^+$He$_{12}$ (icosahedral), K$^+$He$_{10}$ and K$^+$He$_{12}$ (icosahedral). A sudden decrease of the binding energies per atom is found when adding an additional He atom to a snowball of regular structure. This is attributed to the newly-added atom pushing the already bound atoms away from the ion~\cite{Marinetti:2007}. The number of atoms in the first shell is found as N$_1$=10, 12 and 15 for Li$^+$, Na$^+$ and K$^+$, respectively, in perfect agreement with the work cited above~\cite{Rossi:2004}.

Similar calculations on He nanodroplets doped with cationic alkali dimers Ak$_2^+$He$_\mathrm{N}$ (Ak = Li,\,Na,\,K) were performed by the same group~\cite{Bodo:2006JCP,Marinetti:2007JPCA}. Their results are characterized by the competition between the Ak$_2^+$--He and the He--He potential. For Li$_2^+$ the dimer--He interaction clearly dominates leading to a symmetric structure of all even-numbered clusters Li$_2^+$He$_\mathrm{2N}$ up to 2N=18. In the case of K$_2^+$, on the other hand, the He atoms clearly arrange themselves in an asymmetric fashion, thus forming a helium mesh on one side of the dimer. Analysis of the cluster energetics yields values for the number of He atoms in the first solvation shell of N$_1$=6 for Li$_2^+$ and Na$_2^+$, while in the K$_2^+$ case no sign of a shell closure was found.

The study of the distribution of snowball sizes\linebreak Ak$_{1,2}^+$He$_\mathrm{N}$ allows to check the predictions on the number of He atoms in the solvation shells around the ions. The present paper reports on an experimental study of the formation and stability of snowball complexes upon pho\-to\-ionization of alkali metal impurities doped onto helium nanodroplets using fs laser pulses. In contrast to other atoms or molecules, alkali atoms are weakly bound to the surface of He nanodroplets due to the extremely weak attractive interaction between Ak and He atoms~\cite{Stienkemeier:1996}. Above-threshold laser ionization then leads to massive evaporation of He atoms as well as to the expulsion of the ionized dopant particle, which partly entrains He atoms to form a snowball structure. The experimental conditions for the formation of alkali helium snowball complexes are discussed with respect to the different behaviour of monomer and dimer snowballs, Ak$^+$He$_\mathrm{N}$ and Ak$_2^+$He$_\mathrm{N}$. The experimental abundance spectra are compared to theoretical predictions, which agree with the experiment on the first He shell closure for Cs and Rb snowballs. These measurements extend earlier experiments on the formation and stability of alkali clusters embedded in helium nanodroplets using femtosecond photo ionization~\cite{Droppelmann:2008,Schulz:2004}. Furthermore, they provide important complementary information for the interpretation of femtosecond pump-probe experiments studying the dynamics of alkali-helium exciplex formation~\cite{Mudrich:2008,Droppelmann:2004} and of the propagation of vibrational wave packets in alkali dimers formed on helium nanodroplets~\cite{Claas:2007,Claas:2006}.

\section{Experiment}
In the experiment presented here, snowball complexes are studied by ionizing alkali (Ak) atoms and small clusters of various species attached to helium nanodroplets. The droplet machine employed for this study has already been described in detail elsewhere~\cite{Mudrich:2008,Claas:2006}. Briefly, ultrapure helium gas is expanded through an orifice 10\,\textmu m in diameter to form helium nanodroplets. The expansion conditions (p$_{\mathrm{He}}$=60\,bar, nozzle temperature 13\,K) are chosen within the subcritical expansion regime such that photo-ionization signals recorded both at the mass of the neat Ak$^+$ ions as well as at the snowball masses are maximum. The corresponding average droplet size amounts to about 20,000 atoms. The source chamber is connected to the doping chamber by a Beam Dynamics skimmer, 400\,\textmu m in diameter. The He droplets are doped with Ak atoms by passing through a heated pickup cell before entering the detection volume of a commercial quadrupole mass spectrometer inside a third vacuum chamber. Here, the doped beam is ionized either by laser radiation or electron bombardment.

We use a commercial modelocked fs Ti:Sa laser featuring a FWHM pulse length of 150\,fs. The laser wavelength is selected for highest snowball yield relative to the neat ion Ak$^+$ signal, thus resulting in different sets of wavelength and intensity for the different alkalis. The optimal wavelengths for Na, K, Rb and Cs are found to be 755, 730, 800 and 860\,nm, respectively. Using a lens with a focal length of $f=100$\,mm the peak laser intensity in the interaction volume was about 33, 23, 40 and 32\, GW/cm$^2$, respectively.

Upon photo-ionization, the ions Ak$^+$ desorb from the helium surface, partly entraining He atoms from the droplet to form snowball complexes. We find that the fragmentation of alkali clusters Ak$_\mathrm{N}$ upon photo-ionization into atomic ions Ak$^+$ greatly enhances the formation of snowballs Ak$^+$He$_\mathrm{N}$. 
The clusters are formed by increasing the vapor pressure in the pickup cell, so that the helium droplets consecutively pick up more than a single atom. The temperature of the alkali oven was varied in order to maximize the snowball yield resulting in average loading of about 10 alkali atoms per droplet.
\begin{figure}[t!]
\includegraphics[width=\linewidth]{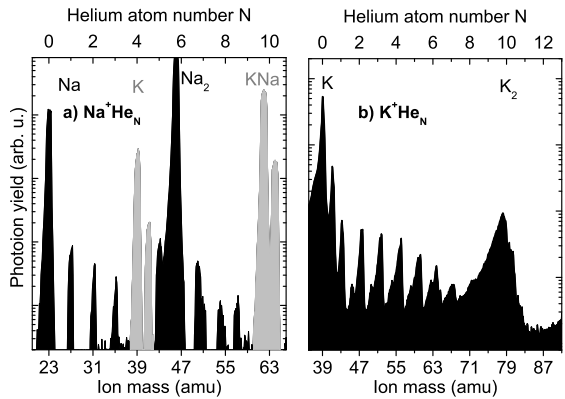}
\caption{\label{MS_leichte}Mass spectra displaying the Ak$^+$He$_\mathrm{N}$ (Ak=Na,\,K) snowball progression on a logarithmic scale. Mass peaks resulting from contamination with other alkali metals or residual water vapour are represented in gray.}
\end{figure}
\section{Results and discussion}
\begin{figure}[t!]
\includegraphics[width=\linewidth]{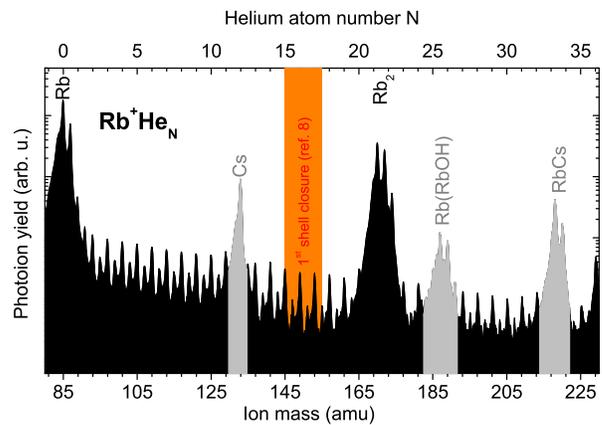}
\caption{\label{MS_Rb}Same as Fig. \ref{MS_leichte}, but for rubidium. The shaded area indicates the expected position of the first shell closure as derived from~\cite{Rossi:2004}. Mass peaks resulting from contamination with cesium or residual water vapour are represented in gray.}
\end{figure}

\begin{figure}[t!]
\centerline{\includegraphics[width=\linewidth]{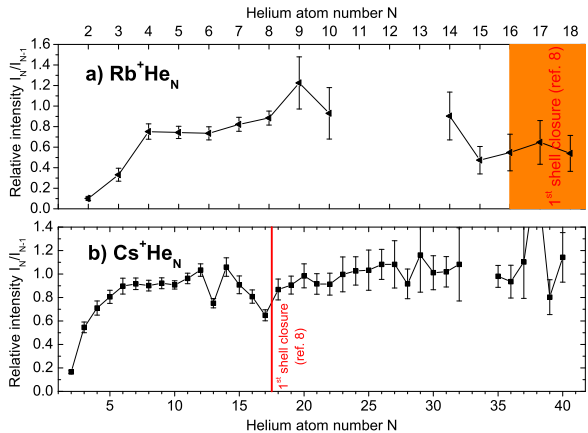}}
\caption{\label{diffMS_zusammen}'Differential mass spectrum' of snowballs showing the relative abundances of neighbouring snowball masses AkHe$_\mathrm{N}$, AkHe$_\mathrm{N-1}$. As in figures~\ref{MS_Rb} and ~\ref{MS_Cs}, note the predicted position of the first shell closure. Some data points are missing due to an overlap of Cs,\,Cs$_2$ peaks with the snowball masses.}
\end{figure}

\begin{figure}[t!]
\includegraphics[width=\linewidth]{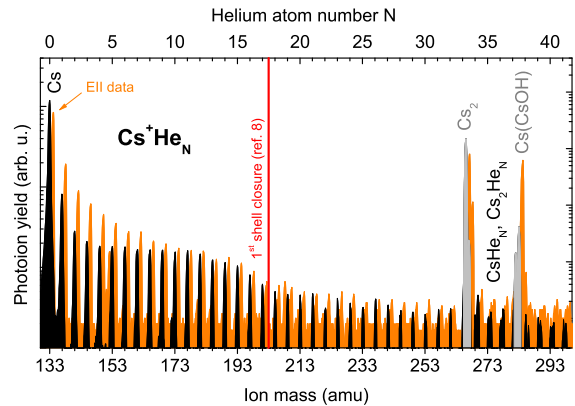}
\caption{\label{MS_Cs}Same as Fig. \ref{MS_leichte}, but for cesium. The Cs$^+$He$_\mathrm{N}$ progression displays a distinct drop in snowball intensity in accordance with the results in~\cite{Rossi:2004}. Orange data were taken by EII and are shifted by 1\,amu for comparison.}
\end{figure}

We systematically study all Ak metals except Li, since the latter's small mass and isotopic composition seriously hamper the evaluation of mass spectrometric data of\linebreak Ak$^+$He$_\mathrm{N}$ snowball complexes. The snowball series recorded in this experiment are depicted in figures~\ref{MS_leichte} to~\ref{MS_Cs}. Snowball formation can clearly be seen for all metals probed. The formation of small snowballs appears to be roughly equally efficient irrespective of the particular Ak species considered: A comparison of the abundance ratios of the snowball Ak$^+$He$_2$ with respect to the ion Ak$^+$ yields values between 0.5\,\% and 1.0\,\% for all shown mass spectra. The data obtained for Na, shown in Fig.~\ref{MS_leichte} a), clearly reveal the presence of small snowballs, Na$^+$He$_{1...3}$. Unfortunately the sample contains some percentage of K, which --- given a count rate for snowball formation of 0.5\,\% of the neat ion --- is enough to superimpose a strong K mass peak onto the snowball series at the mass of Na$^+$He$_4$. All subsequent snowball peaks Na$^+$He$_{\mathrm{N}>4}$ therefore coincide with the peaks of K snowballs K$^+$He$_\mathrm{N}$. Nonetheless we are able to rule out the existence of snowballs larger than Na$^+$He$_{10}$ under our experimental conditions.

K behaves very similarly. The mass spectrum displayed in Fig.~\ref{MS_leichte} b) shows a snowball progression starting with the monomer and eventually disappearing in the dimer peak. Since the data displayed in figures~\ref{MS_leichte} and~\ref{MS_Rb} suffer from low resolution for technical reasons it is not possible to determine the exact end of the snowball series. Nevertheless we can follow the peaks up to K$^+$He$_8$ and can definitely rule out the presence of any snowballs larger than K$^+$He$_{10}$.

In contrast to K atoms, the heavy alkalis produce large snowballs with at least 41 helium atoms attached to the metal ion. Data obtained when doping Rb atoms are depicted in Fig.~\ref{MS_Rb}. The shaded peaks originate from Cs contamination of the sample and residual water vapor in the vacuum chamber. One can follow the snowball series up to Rb$^+$He$_{41}$; due to the isotopical broadening of the Rb$_3^+$ signal we can not unambigously assign further mass peaks to the monomer or the trimer progression Rb$^+$He$_{\mathrm{N}>41}$ or Rb$_3^+$He$_\mathrm{N}$, respectively. Nonetheless, the data clearly cover the mass range, in which the closure of the first He shell around the ionic core is expected according to variational Monte Carlo simulations~\cite{Rossi:2004}. Although Rossi \textit{et al.} do not explicitly treat Rb$^+$He$_\mathrm{N}$ snowballs, their data show an increasing snowball size with Ak atom radius. Therefore, for Rb$^+$ we expect a value N$_1$ for the number of He atoms inside the first solvation shell between the value for K$^+$, N$_1$=15, and the one for Cs$^+$, N$_1$=17.5 He atoms. This range is indicated as a shaded area in Fig.~\ref{MS_Rb}. It is difficult to see clear signs of a shell closure from the data presented in this figure, as the snowball distribution evolves smoothly through the region of interest.

To unambiguously ascertain the position of shell closure from our data, we plot the same data as 'differential mass spectrum' in Fig.~\ref{diffMS_zusammen} a). In this representation we examine the abundance ratio of neighboring snowball mass intensities I$_\mathrm{N}$/I$_\mathrm{N-1}$, so this plot reveals possible shell closures as dips: Supposing an exceptionally abundant snowball Ak$^+$He$_\mathrm{m}$, we expect I$_\mathrm{N}$/I$_\mathrm{N-1}$ to drop significantly below unity for N=m+1. The data shown in this figure is the average over different measurements at various oven and nozzle temperatures to rule out a possible influence of droplet or Ak cluster size on the snowball structure. The differential mass spectrum shows a sudden drop for N=15, followed by a plateau. Compared to the simulations cited above the position of this drop almost coincides with the calculated range of N$_1$=16--18.5.

Fig.~\ref{MS_Cs} features the mass spectrum obtained from Cs-doped He droplets. As in the case of Rb$^+$, we can follow the snowball distribution up to at least Cs$^+$He$_{41}$. However, there might well be more snowballs beyond the scan range employed for this measurement. In addition to the snowball series extending from the Cs$^+$ monomer mass, we see clear evidence for snowballs formed around the Cs dimer ion, Cs$_2^+$He$_\mathrm{N}$. The position of the closure of the first shell, N$_1$=17.5 He atoms, as predicted in Ref.~\cite{Rossi:2004}, is also represented in the graph. In this mass range a step in the abundance spectrum of Cs$^+$He$_\mathrm{N}$ is clearly visible.

To make sure the snowball distribution is not influenced by the ionization scheme employed in this measurement, additional mass spectra are taken by electron impact ionization. The resulting spectrum taken at an impact energy of 70\,eV is shown in Fig.~\ref{MS_Cs} as orange spectrum, shifted by 1\,amu towards higher masses. It shows a smooth decay of small snowball sizes Cs$^+$He$_{\mathrm{N}<10}$ in contrast to the plateau found in the PI mass spectrum. More relevant, however, is the occurrence of the same step at Cs$^+$He$_{15...20}$ that was already found by PI. This shows that the step is indeed a feature intrinsic to the Cs$^+$ ions independent of the ionization mechanism.

The differential mass spectrum reveals another feature, that is easily overlooked in Fig. \ref{MS_Cs}: We note a sudden drop in intensity after Cs$^+$He$_{12}$, hinting at another relatively stable structure in addition to Cs$^+$He$_{16}$. By means of Monte Carlo simulations of alkali-ion doped helium clusters, strong variations of the single-atom evaporative energy were found when adding an additional He atom to an ensemble of regularly arranged solvent atoms like the icosahedral K$^+$He$_{12}$~\cite{Marinetti:2007}. These variations are explained by the large difference between the structures of the regular snowball Ak$^+$He$_\mathrm{N}$ and the subsequent cluster\linebreak Ak$^+$He$_\mathrm{N+1}$: the additional He atom pushes out the regularly bound solvent atoms to a position further away from the ion, thus decreasing their binding energy. 
Such a relatively weakly bound cluster is more susceptible to dissociation. Unfortunately there are no calculations available for Cs$^+$-doped helium clusters. However, it seems plausible to assume a similar behaviour for the insertion of a He atom into Cs$^+$He$_{12}$, as seen in the mass spectrum in Fig.~\ref{diffMS_zusammen} b) for Cs$^+$He$_{13}$.

It is worth mentioning that the spectra of Na and Cs also show dimer snowballs, Ak$_2^+$He$_\mathrm{N}$, while we are unable to observe them for K and Rb. The Na progression Na$_2^+$He$_\mathrm{N}$ can be traced to N=3 before disappearing in the mass peaks of KNa$^+$He$_\mathrm{N}$. However, since our study does not focus on snowballs around dimer ions Ak$_2^+$He$_\mathrm{N}$, they are strongly superimposed in the mass spectra by snowballs formed around monomer ions Ak$^+$ and impurities. Besides, the variations of the evaporation energy between neighbouring clusters are rather weak (ca. 25\,cm$^{-1}$ for Na$_2^+$He$_6$ as compared to Na$_2^+$He$_7$)~\cite{Marinetti:2007JPCA}. The combination of these two facts might well account for the lack of a distinct step in our snowball distribution. The same holds for our measurements of Cs$_2$-doped snowballs, where the snowballs extend at least up to Cs$_2^+$He$_8$. Note that we cannot rule out the existence of larger snowballs, since they would exceed the mass range covered by our measurement.

The formation of snowballs Ak$_2^+$He$_\mathrm{N}$ was found to be governed by the relative strengths of the Ak$_2^+$--He versus the He--He interaction~\cite{Marinetti:2007JPCA}. Since the Ak$_2^+$--He well is deepest for Li and substantially flattens out with increasing atomic number~\cite{Bodo:2006}, one could expect a monotonic change of snowball abundances when increasing the alkali dimer mass from Na$_2^+$ to Cs$_2^+$. However, this trend is probably masked by the effect of varying amounts of internal excitation carried by the dimer ions Ak$_2^+$ of different alkali species. Since a large fraction of the dimer ions Ak$_2^+$ originate from fragmentation of larger clusters, these Ak$_2^+$ may be produced in highly excited vibrational states. Thus, He atoms attached to Ak$_2^+$ are evaporated to leave behind neat Ak$_2^+$ dimer ions, as previously observed by EII of rare-gas clusters in He droplets~\cite{Lewerenz:1995}. Unfortunately, it is beyond the scope of this investigation to assess the amount of energy carried by the individual alkali dimer ions.


\begin{figure}[t!]
\centerline{\includegraphics[width=\linewidth]{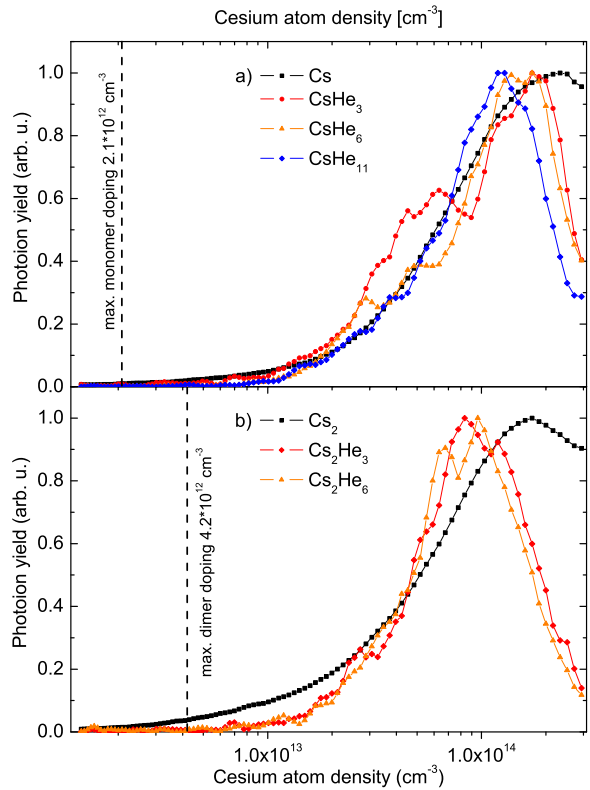}}
\caption{\label{Poisson}Dependence of the yield of different snowballs on the amount of atoms doped into the droplets. Panel a): monomer snowballs Cs$^+$He$_\mathrm{N}$ (N=0, 3, 6, 11), panel b): dimer snowballs Cs$_2^+$He$_\mathrm{N}$ (N=0, 3, 6). The gray dashed line indicates the atom density corresponding to most efficient doping of a single Cs atom. Note the logarithmic scale.}
\end{figure}

The nonresonant nature of the ionization process is expected to enhance fragmentation of the Cs clusters. This assumption is backed by PI spectra taken on snowball masses, which commonly show broad features; e.g., in the case of Cs$^+$He$_7$ we observe a broad maximum peaking around 755\,nm with a FWHM of about 130\,nm. The absence of the atomic Cs D line in the spectra indicates that even single-Cs snowballs Cs$_1^+$He$_\mathrm{N}$ originate predominantly from Cs clusters. To assess the degree of fragmentation occurring in our experiment, doping curves are recorded by varying the oven temperature and simultaneously recording Cs$_\mathrm{M}^+$He$_\mathrm{N}$ mass spectra. Typical examples of these curves are displayed in Fig.~\ref{Poisson}, with panels a) and b) showing the doping dependency of monomer snowballs Cs$^+$He$_\mathrm{N}$ and dimer snowballs Cs$_2^+$He$_\mathrm{N}$, respectively. Within the plots the squares represent neat atoms or dimers Cs$_{1,2}^+$, while the circles, triangles and diamonds represent snowballs Cs$_{1,2}^+$He$_\mathrm{N}$. Taking into account the atom density for maximum doping of a single atom of $2.1\cdot10^{12}$\,cm$^{-3}$ (indicated as a dashed line), we deduce that the vast majority of detected Cs$_{1,2}^+$ ions originate from fragments of clusters Cs$_\mathrm{N}$ formed on the droplets. Similar behaviour has been reported for the case of silver clusters~\cite{Doeppner:2007}.

It is interesting to note that the dependence of snowball abundance on the doping level differs from the one of neat Cs atoms: At Cs vapor pressures below $3\cdot10^{12}$\,cm$^{-3}$ the curvature of the different doping curves increases with increasing N, indicating that bigger snowballs originate from bigger neat Cs clusters. However, further increasing the doping level above ca. 10$^{14}$\,cm$^{-3}$ leads to falling snowball abundances before reaching the maximum in neat Cs intensity. This decline in the production of snowballs with increasing alkali doping is related to the influence of doping on the helium nanodroplet size: During the cluster growth each consecutive Cs atom dissipates binding energy into the droplet. This energy is released by evaporation of helium atoms, leading to a reduced number of helium atoms available for snowball formation.

This effect is more pronounced when photo-ionizing dimer snowballs. The abundance of dimer snowballs\linebreak Cs$_2^+$He$_\mathrm{N}$ (Fig.~\ref{Poisson} b) passes its maximum at lower doping levels than the monomer snowball signal (Fig.~\ref{Poisson} a). This is caused by the internal degrees of freedom of the ionized dimer Cs$_2^+$. In addition to boiling off He atoms from the droplet upon ionization, the Cs$_2$ bond can carry internal energy
, which is dissipated by evaporative cooling of the snowball Cs$_2^+$He$_\mathrm{N}$ after it has left the droplet~\cite{Lewerenz:1995}. This leads to further reduction of the snowball abundance at high doping levels.

Examining the maximum positions of the neat ions Cs$^+$, Cs$_2^+$, we find the monomer maximum to be shifted towards higher Cs doping than the dimer. This is in accordance with observations of the fragmentation of alkali cluster ions Ak$_\mathrm{N}^+$ with Ak=Na, K. Fragmentation of large ionized clusters, Na$_\mathrm{N}^+$ and K$_\mathrm{N}^+$, predominantly produces monomer fragment ions Ak$^+$ as compared to dimer ions Ak$_2^+$~\cite{Brechignac:1989JCP,Brechignac:1990}.

The doping curves recorded at monomer snowball\linebreak masses Cs$^+$He$_\mathrm{N}$ (circles, triangles and diamonds in Fig. \ref{Poisson} a)) show a shoulder towards lower doping levels, which is very pronounced for small Cs$^+$He$_{\mathrm{N}<4}$, and whose width and intensity gradually fade away when observing larger snowballs up to CsHe$_{20}$. We take this as a sign of massive influx into these snowball masses from individual neat Cs cluster sizes. Based on the position of the peak relative to the position of maximal monomer doping, we roughly assign clusters around Cs$_{25}$ as the most probable precursor.

\section{Conclusion}
In a systematic study of helium snowballs formed around alkali ions we have multiply loaded helium nanodroplets with all alkali metals except Li and created snowballs by fs PI and by EII of alkali clusters. The data for Na and K allow only the study of rather small snowballs of size up to Na$^+$He$_3$ and K$^+$He$_{10}$. The results for the heavier alkali species permit comparison with variational Monte Carlo simulations~\cite{Rossi:2004,Marinetti:2007}. The number of atoms observed in the first helium shell around Rb$^+$ and Cs$^+$ ions is found to agree with theoretical predictions. The differential spectrum for Cs$^+$ hints at an additional subshell closure for Cs$^+$He$_{12}$, possibly due to the swelling of the isocahedral helium shell around the Cs ion to accommodate the next He atom; however, for lack of theoretical data this assumption cannot be verified. Snowball formation around alkali dimer ions is observed for Na$_2^+$ and Cs$_2^+$.

Doping curves reveal strongly increased probability of formings snowballs around Cs$^+$ and Cs$_2^+$ fragment ions coming from fragmenting large Cs clusters. The order of the maxima of doping curves measured for different snowball masses is interpreted in terms of massive evaporation of helium upon cluster formation and ionization.

In future experiments we plan to replace the fs multi-photon ionization scheme by single-photon ionization with nanosecond lasers. This will allow an exact determination (and adjustment) of the energy deposited into the clusters and will thus drastically reduce fragmentation. Furthermore, ionization just above the IP will rule out any contribution from higher excited ionic states. In that way the interpretation of our data will be greatly simplified. Furthermore, experiments with amplified fs pulses may induce much more extended snowball series, as observed with magnesium~\cite{Doeppner:2001}. Fs pump-probe measurements may reveal the presence of caging-effects~\cite{Doeppner:2007} and shed some light onto the issue whether alkali clusters are surface bound or immersed into the He droplets. Furthermore, the theoretical study of snowball formation around dimer ions Ak$_2^+$ for the heavy alkalis Rb and Cs is desirable to interpret their element-specific appearance and stability patterns. 
Besides, modelling the dynamics of fragmentation of alkali clusters Ak$_\mathrm{N}$ on the droplets may add to the understanding of snowball formation.

\section{Acknowledgment}
Financial support by the Deutsche Forschungsgemeinschaft is gratefully acknowledged.


\end{document}